\documentclass[prd,twocolumn,nofootinbib,aps,superscriptaddress,tightenlines,preprintnumbers]{revtex4}
\usepackage{latexsym}
\usepackage{amsmath}
\usepackage{amssymb}
\usepackage{amsfonts}
\usepackage{comment}
\usepackage{graphicx}
\usepackage{verbatim}

\newcommand{\be}{\begin{equation}}
\newcommand{\ee}{\end{equation}}
\newcommand{\bea}{\begin{eqnarray}}
\newcommand{\eea}{\end{eqnarray}}
\newcommand{\mn}{{\mu\nu}}
\newcommand{\rs}{{\rho\sigma}}

\newcommand{\call}{\mathcal{L}}

\newcommand{\pa}{\partial}
\newcommand{\ab}{\alpha\beta}
\newcommand{\tih}{\tilde{h}}

\begin{document}

\preprint{\hbox{CALT 68-2729}  } 

\title{Lorentz Violation in Goldstone Gravity} 

\author{Sean M. Carroll}
\affiliation{California Institute of Technology, Pasadena, CA 91125 }

\author{Heywood Tam}
\affiliation{California Institute of Technology, Pasadena, CA 91125 }

\author{Ingunn Kathrine Wehus}
\affiliation{Department of Physics, University of Oslo, P.O.\ Box 1048
  Blindern, N-0316 Oslo, Norway}
\affiliation{California Institute of Technology, Pasadena, CA 91125 }

\begin{abstract}
We consider a theory of gravity in which a symmetric two-index tensor in Minkowski spacetime acquires a vacuum expectation value (vev) via a potential, thereby breaking Lorentz invariance spontaneously. When the vev breaks all the generators of the Lorentz group, six Goldstone modes emerge, two linear combinations of which have properties that are identical to those of the graviton in general relativity.  Integrating out massive modes yields an infinite number of Lorentz-violating radiative-correction terms in the low-energy effective Lagrangian. We examine a representative subset of these terms and show that they modify the dispersion relation of the two propagating graviton modes such that their phase velocity is direction-dependent. If the phase velocity of the Goldstone gravitons is subluminal, cosmic rays can emit gravi-Cherenkov radiation, and the detection of high-energy cosmic rays can be used to constrain these radiative correction terms. Test particles in the vicinity of the Goldstone gravitons undergo longitudinal oscillations in addition to the usual transverse oscillations as predicted by general relativity. Finally, we discuss the possibility of having vevs that do not break all six generators and examine in detail one such theory.
\end{abstract}

\date\today
\maketitle

\section{Introduction}

The existence of massless particles is conventionally explained by the requirement to preserve gauge symmetries. In the case of electromagnetism, the masslessness of the photon is required so that local $U(1)$ gauge invariance is maintained; in the case of general relativity, the masslessness of the graviton has its origin in diffeomorphism invariance. 

In 1963, Bjorken proposed an alternative viewpoint: the photon can be a Goldstone boson associated with the spontaneous breaking of Lorentz invariance \cite{Nambu:1960xd, Goldstone:1961eq, Goldstone:1962es, Bjorken:1963vg, Bjorken:2001pe}. The idea was subsequently generalized and applied to the case of gravity by Phillips and others \cite{PhysRev.146.966,Ohanian:1970qe,Kraus:2002sa,Berezhiani:2008ue}.

In ordinary Maxwell electrodynamics, gauge invariance reduces the four components of the vector potential $A_\mu$ down to the two propagating degrees of freedom of a massless spin-1 particle.  Gauge invariance forbids a potential $V(A_\mu)$, which keeps the photon massless and prohibits a longitudinal mode, and it also forbids kinetic terms such as $(\partial_\mu A^\mu)^2$, which would  allow a spin-0 mode to propagate.  In the Goldstone approach, there is no gauge invariance, and the vector field acquires a vev via a potential.  Regardless of the form of the vev, there are always three massless Goldstone excitations, all of which would propagate for a generic choice of kinetic term.  To avoid the extra degree of freedom, we can choose the Maxwell kinetic term, even though it is not required by gauge invariance.  Then two linear combinations of the Goldstone modes have exactly the same properties as the photon in electromagnetism. The remaining longitudinal mode is auxiliary, and does not propagate, so that the theory is indistinguishable from electromagnetism in the low energy limit.  (In the presence of Lorentz violation, Goldstone's theorem no longer ensures one propagating mode for each broken symmetry generator.)  This identification can be overturned by radiative corrections, since there is no gauge invariance to protect the form of the propagator.

The graviton case is similar, except that now it is a symmetric two-index tensor that acquires a vev. A propagating massless spin-2 particle has two degrees of freedom. Because the Lorentz group has six generators, there are sufficient degrees of freedom in the Goldstone bosons to reproduce the graviton.  However, we will see that this is not automatic, as in the photon case; whether we get the correct Goldstone modes to recover the transverse-traceless oscillations of conventional gravitons will depend on the choice of vev.

Recently, Kostelecky and Potting examined in detail the scenario in which a symmetric two-index tensor acquires a vev via a potential \cite{Kostelecky:2009zr}. With a kinetic term quadratic in derivatives and preserving diffeomorphism invariance, they found that, just as in the photon case, two linear combinations of the resulting six Goldstone bosons obey the linearized Einstein's equations in a special gauge (which they termed the `cardinal' gauge), while the remaining four linear combinations do not propagate. Together with four additional massive modes, they account for the ten degrees of freedom contained in the two-index symmetric tensor. By requiring self-consistent coupling to the energy-momentum tensor, they also demonstrated that the theory can be used to construct a nonlinear theory via a bootstrap procedure (analogous to the way in which general relativity is obtained from the linearized theory). We expect the massive modes to be near the Planck scale, outside the low-energy theory, so the nonlinear theory is equivalent to general relativity with conventional coupling to matter.  

Kraus and Tomboulis \cite{Kraus:2002sa} pointed out that these massive modes nevertheless have a crucial effect:  integrating them out introduces an infinite number of radiative-correction terms to the low-energy Lagrangian, which can change the theory in important ways.  
Since these corrections are not controlled by gauge invariance, in general they will modify the dispersion relations of the Goldstone modes. At higher order, therefore, the Goldstone bosons arising from Lorentz violation can, in principle, be distinguished from the graviton in linearized general relativity. 

In this paper, we examine some of these correction terms and study their effects on the properties of the Goldstone bosons.  (The terms we consider are those that are most straightforward to analyze, but their impacts should be generic.)  We find that, for a general vev, these terms modify the dispersion relations of the Goldstone modes in such a way that their speed of propagation is anisotropic. If the speed is subluminal in some directions, gravi-Cherenkov radiation by cosmic rays becomes possible. Observations of high-energy cosmic rays thus allow us to constrain these higher-order radiative corrections. These corrections also effect the polarization tensors of the conventional gravitons, leading to longitudinal oscillations in the motion of test particles, in addition to the conventional transverse $+$ and $\times$ patterns predicted in general relativity.  This could lead to novel experimental tests of the theory, although we do not know of any constraints on this phenomenon from currently available data.

Another difference between Goldstone gravity and general relativity is that the former predicts the existence of other massless particles in addition to the two conventional massless spin-2 polarizations. This is reminiscent of the photon case, in which a longitudinal mode (in addition to the two transverse modes) becomes dynamical in the presence of the radiative corrections induced by integrating out the massive modes. Analogously, we expect that there should be four longitudinal Goldstone bosons that can become dynamical. The polarization tensors of these modes can be written as a sum of eight symmetric tensors constructed from $k_\mu$ and the vev. By imposing the four cardinal gauge conditions, we can relate these eight coefficients, leaving four independent parameters for the four Goldstone modes.  

In the next section, we briefly review the case of Goldstone photons, including the effects of radiative corrections as emphasized in \cite{Kraus:2002sa}.  We then carry out an analogous analysis for gravitons, showing how radiative corrections bring to live new massless modes.  In Section IV we concentrate on the two modes of the graviton, demonstrating that they propagate anisotropically in the presence of a generic vev and considering some experimental limits on the corresponding parameters.  In Section V we examine models where the vev doesn't completely break the Lorentz group, and gravitons are only partially constructed from Goldstone bosons, or they originate from residual diffeomorphism invariance.  A series of Appendices describes the relationship between different patterns of symmetry breaking and the modes corresponding to gravitons.

\section{Goldstone Electromagnetism}

\subsection{Photons as Goldstone Bosons}

Before we delve into the graviton case, we first discuss the scenario in which the photon arises as a Goldstone boson of spontaneous Lorentz violation, commonly known as the `Bumblebee' model \cite{Bluhm:2004ep}. We will see below that the graviton case mirrors the vector case. 

We consider the Lagrangian for a vector field $A_\mu$,
\be \label{emlagrangian}
\call = -\frac{1}{4}f_\mn f^\mn - V(\bar{A}_\mu, a_\mu),
\ee
where $A_\mu = \bar{A}_\mu + a_\mu$ and $f_\mn = \pa_\mu A_\nu - \pa_\nu A_\mu = \pa_\mu a_\nu - \pa_\nu a_\mu$ is the corresponding field-strength tensor. The potential $V$ gives $A_\mu$ a vev $\bar{A}_\mu$ (with $\partial_\mu \bar{A}_\nu = 0$), thereby violating Lorentz invariance spontaneously. For a thorough analysis of the case for which $\bar{A}_\mu$ is spacelike, see \cite{Kraus:2002sa}. 

We consider here the usual Maxwell kinetic term, which by itself
preserves gauge invariance, as our aim is to have a theory that
reproduces electromagnetism at lowest order.  The stability of theories with more generic kinetic terms was considered in \cite{Carroll:2009em}. 

The Goldstone boson fields can be constructed from the vev by the action of spacetime-dependent infinitesimal Lorentz transformations,
\be \label{photon}
a_\mu = -{\Theta_\mu}^\nu(x) \bar{A}_\nu.
\ee
Here, $\Theta_\mn$ is an antisymmetric tensor of the form
\be \label{eptensor}
\left( \begin{array}{cccc} 
0 & \beta_1 & \beta_2 & \beta_3 \\
-\beta_1 & 0 & \theta_3 & -\theta_2 \\
-\beta_2 & -\theta_3 & 0 & \theta_1 \\
-\beta_3 & \theta_2 & -\theta_1 & 0 \end{array}\right),
\ee
where $\beta_i = \bar{\beta}_i e^{ik_\alpha x^\alpha}$ are the infinitesimal rapidities corresponding to boosts, and $\theta_i = \bar{\theta} e^{ik_\alpha x^\alpha}$ are the infinitesimal angles corresponding to rotations. Note that the three Goldstone modes $a_\mu$ are orthogonal to the vev $\bar{A}_\mu$. The remaining length-changing mode (parallel to $\bar{A}_\mu$) is massive.  

We can consider vevs $\bar{A}_\mu$ that are timelike or spacelike. When it is timelike, without loss of generality we can boost to a frame in which only $A_0 \neq 0$. This breaks the original $SO(3,1)$ to $SO(3)$, preserving rotational invariance. From Eq.\eqref{photon}, the three Goldstone bosons come from the three broken boost generators, and are given by
\bea
a_\mu &=& -\Theta_{\mu 0}
\\
&=& \left( \begin{array}{c} 
0 \\
\beta_1 \\
\beta_2 \\
\beta_3 \end{array}\right). 
\eea 

Each choice of the vev corresponds to a particular gauge in electromagnetism. Having a timelike vev is equivalent to the Coulomb gauge, in which we set the scalar potential to zero ($A_0=0$).  That is, the physics of the theory is completely equivalent to that of free Maxwell electrodynamics, but with a particular gauge condition imposed.
This gauge choice is compatible with the transverse condition ($k_\mu A^\mu = 0$) that we usually impose in electromagnetism. Together these are consistent with the Lorenz gauge, making a timelike vev a natural gauge choice to describe a free photon. For example, if we want to describe a photon moving in the $x_i$ direction, we can just set $A_i$ to zero.  

If instead $\bar{A}_\mu$ is spacelike, we can rotate axes such that only $\bar{A}_3 \neq 0$. This reduces the $SO(3,1)$ symmetry that we begin with to $SO(2,1)$, resulting in three Goldstone modes (one boost and two rotations):
\bea
a_\mu &=& -\Theta_{\mu 3}
\\
&=& \left( \begin{array}{c} 
\beta_3 \\
\theta_2 \\
-\theta_1 \\
0 \end{array}\right). 
\eea 

Having a spacelike vev is equivalent to imposing the axial gauge
($\vec{s}\cdot\vec{a} = 0$), where $\vec{s}$ is a unit spatial vector.
In order to describe a photon that propagates in a direction orthogonal to $\bar{A}_\mu$, $a_\mu$ is necessarily unbounded somewhere at spatial infinity. There is thus a question whether the Lorentz-violating theory, as an effective field theory, is capable of describing a photon in the axial gauge. Since the field value can be large, we should, in the spirit of effective field theory, retain higher-order kinetic terms in the Lagrangian.  We won't pursue this issue in this paper.
 
\subsection{Radiative Corrections and Dispersion Relations of the Goldstone Modes}

As we have seen, the vev $\bar{A}_\mu$ always leads to three Goldstone bosons, which can be classified into two transverse modes and one longitudinal mode. The transverse modes satisfy the condition $k^\mu a_\mu = 0$. With the kinetic term in \eqref{emlagrangian}, they satisfy the dispersion relation $k^\mu k_\mu = 0$, and thus propagate isotropically at the speed of light. Hence, they have the right properties to be identified as the photon. 

The remaining longitudinal degree of freedom is orthogonal to the two transverse modes. This allows us to specify its polarization as
\be \label{longmode}
\epsilon^{(longitudinal)}_\mu = k_\mu -	 \frac{(\bar{A}^\alpha k_\alpha)}{A^\beta A_\beta}\bar{A}_\mu.
\ee
At lowest order, this longitudinal mode does not propagate, and corresponds to the pure-gauge mode in electromagnetism.  

As we will see later, this way of decomposing the Goldstone modes into transverse and longitudinal degrees of freedom will be highly similar in the graviton case. Expressing the longitudinal mode in the basis $k_\mu$ and $\bar{A}_\mu$ makes it automatically orthogonal to the transverse modes. 

As was pointed out in \cite{Kraus:2002sa}, we expect that there would
be higher-order radiative correction terms induced in the low-energy
effective Lagrangian as we integrate out the massive fluctuations of
$A_\mu$. These terms will in general modify the dispersion relations
of the Goldstone bosons. If we restrict our attention to only two
derivatives, there are seven such terms, which are listed in
\cite{Kraus:2002sa} and which take the form:
\bea
&&f_1(A^2)\partial_\mu A_\nu \partial^\mu A^\nu \notag\\
&&f_2(A^2)\partial_\mu A_\nu \partial^\nu A^\mu \notag\\
&&f_3(A^2)A^\mu A^\alpha \partial_\mu A_\nu \partial_\alpha A^\nu \notag\\
&&f_4(A^2)A^\nu A^\alpha \partial_\mu A_\nu \partial_\alpha A^\mu \label{radcorr}\\
&&f_5(A^2)A^\nu A^\alpha \partial_\mu A_\nu \partial^\mu A_\alpha \notag\\
&&f_6(A^2)A^\mu A^\nu A^\alpha \partial_\mu \partial_\nu A_\alpha \notag\\
&&f_7(A^2)A^\mu A^\nu A^\alpha A^\beta \partial_\mu A_\nu \notag
\partial_\alpha A_\beta \,,
\eea
where $f_i(A^2)$ are scalar functions of $A^\mu A_\mu$. This list exhausts all possible such terms, since $A^\mu A_\mu$ is a constant. The situation will be different in the two-index case, where infinitely many such terms can be generated in the effective Lagrangian, as we will discuss later.

If we assume that $\bar{A}_\mu a^\mu$ is small, the first three terms in (\ref{radcorr}) dominate over the rest. They modify the dispersion relations of the two transverse Goldstone bosons to 
\be \label{photontrans}
(1+d_1)k^\mu k_\mu + d_2 (\bar{A}^\mu k_\mu)^2 = 0,
\ee
where $d_1$ and $d_2$ are undetermined coefficients and are presumably small. The additional term implies that the phase velocity of the two transverse modes is anisotropic. 

Meanwhile, in the presence of these radiative corrections, the longitudinal mode becomes dynamical and has the dispersion relation
\be
k^\mu k_\mu + d_3(\bar{A}_\mu k^\mu)^2 = 0,
\ee
where $d_3$ is an undetermined coefficient. 

\section{Goldstone Gravity}

\subsection{Gravitons as Goldstone Bosons}

The analysis of spontaneous Lorentz violation via a symmetric two-index tensor is in many ways similar to the vector case that we previously discussed. In particular, we will focus on a model called `cardinal gravity', introduced recently by Kostelecky and Potting \cite{Kostelecky:2009zr}. They showed that when a two-index symmetric tensor acquires a vev which breaks all six generators of the Lorentz group in Minkowski spacetime, two linear combinations of the resulting Goldstone modes have properties that are identical to those of the graviton in a special (cardinal) gauge in linearized general relativity.  We have included our own version of this argument in Appendix B. 

As in the photon case, higher-order radiative correction terms resembling \eqref{radcorr} will generically appear in the low-energy effective Lagrangian as we integrate out the four massive modes to extract their contribution to the low energy physics. In the two-index case, there are infinitely many such terms. In this paper, we will focus on a representative subset of these terms, and examine their resulting Lorentz-violating effects on the Goldstone modes. For example, in the presence of these higher-order terms, two linear combinations of the six Goldstone modes that are to be identified as the graviton will now propagate at different phase velocities in different directions. In addition, the four remaining linear combinations that are originally auxiliary will now become dynamical, just like the longitudinal mode in the vector case.  

We begin with the Lagrangian
\bea\nonumber \label{grlagrangian}
\call &=& \frac{1}{2}[(\pa_\mu \tih^{\mn})(\pa_\nu \tih)-(\pa_\mu \tih^{\rs})(\pa_\rho {\tih^\mu}_\sigma)
\\ \nonumber
&& + \frac{1}{2}\eta^{\mn}(\pa_\mu \tih^{\rs})(\pa_\nu \tih_{\rs}) -\frac{1}{2}\eta^{\mn}(\pa_\mu \tih)(\pa_\nu \tih)] 
\\
&& + \mbox{(radiative corrections)}- V(\tih_{\mn}\tih^{\mn}),
\eea
where $\tih^{\mn}$ is a symmetric two-index tensor field defined on a spacetime with Minkowski metric $\eta_{\mn}$.  In analogy to the electromagnetic case, we have chosen
the linearized Einstein-Hilbert kinetic term, which by itself preserves diffeomorphism invariance
($\tih_\mn \rightarrow \tih_\mn + \partial_{(\mu}\xi_{\nu)}$, for some vector $\xi^\mu$). 

As in the vector case, the field $\tih_\mn$ acquires a vev $H_\mn$ via the potential $V$. The Goldstone modes that result are given by acting spacetime-dependent infinitesimal Lorentz transformations on this vev \cite{Bluhm:2004ep,Bluhm:2006im}:
\be \label{goldstone}
h_{\mn} = -{\Theta_\mu}^\alpha H_{\alpha\nu} - {\Theta_\nu}^\alpha H_{\mu\alpha},
\ee
where $\tih_\mn = H_\mn + h_\mn$ and $\Theta_{\mn}$ is as defined in \eqref{eptensor}. Unless stated otherwise, from now on we assume that $H_\mn$ breaks all six generators of the Lorentz group, and thus gives rise to six potential Goldstone bosons. 
 
Note that in the form of \eqref{goldstone}, the Goldstone bosons
automatically fulfill four conditions, dubbed `cardinal' by Kostelecky
and Potting in \cite{Kostelecky:2009zr}:
\bea \label{gaugecond}
\eta^{\mn}\epsilon_\mn &=& 0 \\ 
{H^\mn} \epsilon_{\mn} &=& 0 \\
{H^\mu}_\alpha H^{\nu\alpha} \epsilon_{\mn} &=& 0 \\ \label{gaugecond4}
H^{\mu\alpha}H_{\alpha\beta}H^{\beta\nu} \epsilon_{\mn} &=& 0,
\eea
where $h_\mn = \epsilon_\mn e^{ik_\alpha x^\alpha}$.  Since we could diagonalize
$H_\mn$ via an appropriate orthogonal transformation, there can be at most four such
independent constraints, one for each eigenvalue.  Contracting $\epsilon_\mn$
with terms of higher order in $H_\mn$ (eg. $H^{\mu\alpha}
H_{\alpha\beta} H^{\beta\gamma} H_{\gamma\nu}$) also
yields zero, but the resulting constraints are not independent.

The cardinal conditions are very similar to that ($\bar{A}_\mu a^\mu = 0$) in the vector case, but now there are four orthogonality conditions instead of one. They can be viewed as `directions' along which the massive modes reside (just as the length-changing mode of the vector is parallel to the vev). There are thus in general four massive degrees of freedom in the theory.  

Kostelecky and Potting demonstrated that the cardinal gauge is attainable
for generic $k^\mu$ in general relativity. In Appendix B we derive necessary and sufficient conditions under which the cardinal gauge is a valid gauge choice. 

Starting with the ten independent components in $h_\mn$, imposing the four cardinal gauge conditions reduces that to six, which is exactly the right number to accomodate the six Goldstone modes. The situation becomes more complicated when the vev does not break all six generators. In that case, there are fewer Goldstone bosons, as well as fewer gauge conditions. However, there might also be residual diffeomorphism invariance. The theory can contain massless excitations that originate from spontaneous Lorentz violation and/or diffeomorphism invariance.

As in the photon case, it is most convenient to decompose the six Goldstone modes into two linear combinations that are transverse, and four other orthogonal linear combinations. The two transverse modes obey the linearized Einstein's equations and have the dispersion relation 
\be
k^\mu k_\mu = 0,
\ee
corresponding to masseless particles propagating isotropically at the speed of light.
These can therefore be identified as the graviton. 
The remaining four modes are auxiliary and do not propagate. At this order, the theory is thus equivalent to linearized general relativity in the cardinal gauge, if we treat the massive modes as absent at low energies.

\subsection{Radiative Corrections and Dispersion Relations}

Corrections to the effective field theory arise from integrating out the massive modes.
As in the photon case, the resulting radiative-correction terms induce additional Lorentz-violating effects when $\tilde{h}_\mn$ acquires a vev. As before, we restrict our attention to only terms that are quadratic in derivatives of $h_\mn$. We will demonstrate that these terms will modify the dispersion relations of the two transverse linear combinations that correspond to the graviton. We also argue that, just as the longitudinal mode in the vector case, the four remaining Goldstone modes become dynamical. 

There are four types of kinetic terms that are independent of $H_\mn$. The terms and their corresponding contributions to the equation of motion are as follows: 
\bea \label{term1}
\pa_\rho h_\mn \pa^\rho h^\mn &\rightarrow& 2\Box h_\mn \\ \label{term2}
\pa_\mu h^\mn \pa_\nu h &\rightarrow& \pa_\mu \pa_\nu h + \eta_\mn \pa_\rho \pa_\sigma h^\rs \\ \label{term3}
\pa_\mu h^\rs \pa_\rho {h^\mu}_\sigma &\rightarrow& 2\pa_{(\mu|} \pa_\sigma {h^\sigma}_{|\nu)} \\ \label{term4}
\pa_\mu h \pa^\mu h &\rightarrow& 2\eta_\mn \Box h.
\eea
Each of these terms already appears in the Lagrangian (\ref{grlagrangian}), with specific numerical coefficients.  The corrections will change the value of these coefficients, generically leading to violations of diffeomorphism invariance. 

At linear order in $H_\mn$ we have the following possible kinetic terms, and their
contributions to the equation of motion: 
\bea
H^{\ab}\pa_\alpha h_{\rs}\pa_\beta h^{\rs} &\rightarrow& 2H^{\ab}\pa_\alpha \pa_\beta h_{\mn}
\\
H^{\ab}\pa^\rho h_{\alpha\rho} \pa^\sigma h_{\beta\sigma} &\rightarrow& 2{H_{(\mu|}}^\beta\pa_{|\nu)}\pa^{\sigma} h_{\beta\sigma}
\\ \nonumber
H^{\ab}\pa_\sigma h_{\ab} \pa_\rho h^{\rs} &\rightarrow& H_{\mn}\pa^\alpha \pa^{\beta} h_{\ab} 
\\ \label{termc}
&& + H^{\ab} \pa_\mu \pa_\nu h_{\ab}
\\ \label{termd}
{H^\alpha}_\beta\pa_\rho {h_{\alpha\sigma}} \pa^\rho h^{\beta\sigma} &\rightarrow& 2{H_{(\mu|}}^\sigma \Box h_{\sigma|\nu)}
\\
H^{\ab}\pa_\rho {h_{\ab}} \pa^\rho h &\rightarrow& H_{\mn}\Box h + (H^{\ab}\Box {h_{\alpha\beta}})\eta_{\mn}
\\ \nonumber
H^{\ab}\pa_\alpha h_{\beta\sigma} \pa_\rho h^{\rho\sigma} &\rightarrow& {H_{(\mu|}}^{\alpha}\pa_\alpha \pa^\beta h_{\beta|\nu)} 
\\ \label{termf}
&& + H^{\ab}\pa_{(\mu|} \pa_\alpha h_{\beta|\nu)}
\\ \nonumber
H^{\ab}\pa_\alpha {h_{\beta\rho}} \pa^\rho h &\rightarrow& {H_{(\mu|}}^\alpha\pa_\alpha \pa_{|\nu)} h  
\\
&& + (H^{\ab}\pa^\sigma \pa_\alpha h_{\beta\sigma})\eta_{\mn}
\\
H^{\ab}\pa_\alpha h \pa_\beta h &\rightarrow& 2H^{\ab} \pa_\alpha \pa_\beta h \eta_{\mn}.
\eea
Unlike the photon case, there are infinitely many radiative correction terms that can be generated at higher orders in the vev. Assuming that $H_\mn$ is in general small compared to the background metric, we will focus only on those that either do not depend on, or those linear in, $H_\mn$. We will later discuss a possible experimental test to constrain $H_\mn$.

We first consider the four auxiliary modes. In the form of \eqref{goldstone}, they obey the four cardinal gauge conditions \eqref{gaugecond}. They are also orthogonal to the two transverse degrees of freedom that correspond to the graviton,
\be \label{orthocond}
\epsilon^{(aux)}_{\mn} \epsilon^{\mn}_{(trans)} = 0.
\ee
Together, these are six conditions that reduce the ten independent components of $\epsilon^{(aux)}_\mn$ to four degrees of freedom. In analogy to \eqref{longmode} in the photon case, we can express these four modes in terms of the wave vector and the vev as
\bea \nonumber
\epsilon^{(aux)}_\mn &=& b_1 \eta_\mn + b_2 H_\mn + b_3 {H_\mu}^\alpha H_{\alpha\nu} \\ \nonumber
&&+ b_4 H_{\mu\alpha} H^{\alpha\beta} H_{\beta\nu} + b_5 k_\mu k_\nu \\ \nonumber
&&+ b_6 H_{(\mu|\alpha}k^\alpha k_{|\nu)} + b_7 H_{(\mu|\alpha} H^{\alpha\beta}k_\beta k_{|\nu)} \\ 
&&+ b_8 H_{(\mu|\alpha} H^{\alpha\beta} H_{\beta\gamma}k^\gamma k_{|\nu)},
\eea
where the eight coefficients $b_i$ are constrained by imposing the four cardinal gauge conditions \eqref{gaugecond} - \eqref{gaugecond4}. This leaves four independent coefficients for the four modes. 

The basis polarization tensors $\epsilon^{(aux)}_\mn$ are chosen so that the conditions \eqref{orthocond} are automatically satisfied. At lowest order, these four modes do not propagate (as is demonstrated in Appendix B). However, in the presence of the radiative correction terms, we expect that they become dynamical, similar to the longitudinal mode in the vector case. There will now be a contribution from \eqref{term1}, which adds the term $k^\mu k_\mu$ to their dispersion relation. We do not pursue the calculation of the dispersion relations of these auxiliary modes in this paper. The method to do so can be found in \cite{Kraus:2002sa}, in which the dispersion of the longitudinal mode in the photon case is computed.

\section{Anisotropic Propagation}

Now we consider the effects of the radiative correction terms on the two transverse propagating linear combinations, which will be the main focus of this paper. We will not be considering all of the terms, however, as the task of diagonalizing the resulting equations of motion is highly nontrivial. Rather, we  focus on a number of representative terms and see what are some of the Lorentz-violating effects typical in this theory. This will provide a guide on how we can experimentally differentiate the theory from general relativity, given that the two theories are identical at lowest order. 

\subsection{Dispersion Relations}

Of the four terms \eqref{term1} $\rightarrow$ \eqref{term4}, only the first term modifies the dispersion relation, which becomes
\be
(1+c_1)k^\mu k_\mu = 0,
\ee
where $c_1$ is some undetermined constant. In the absence of other terms in the dispersion relation, this correction is immaterial. We can divide by $1+c_1$ and obtain the usual $k^\mu k_\mu = 0$, so excitations propagate isotropically along the light cones.

If we also incorporate the radiative corrections that are linear in $H_\mn$, the equations of motion are still easily diagonalizable except for Eq.\eqref{termd} and \eqref{termf}. We will thus focus on the effects of the other six terms.  The polarization tensors of the transverse Goldstone modes remain unchanged, but their dispersion relations are now modified:  
\bea \label{moddis}
k_\mu k^\mu - c_2 H^{\mn}k_\mu k_\nu = 0,
\eea
where $c_2$ is some undetermined coefficient that is expected to be small. 

As in \eqref{photontrans}, the effect of the additional term in the dispersion relation is to make the phase velocity of the transverse modes become anisotropic for a generic vev. The phase velocity is given by the ratio of the frequency $\omega$ and magnitude of the momentum $k$,
\be
v = \frac{\omega}{|\vec{k}|} = 1 - \frac{c_2}{2}n_\mu H^{\mn} n_\nu,
\ee
where $n_\mu = (1,\vec{n})$ and $\vec{n} = \vec{k}/{|\vec{k}|}$.

Note that in the case where $H_{\mn}$ can be written as $t_\mu t_\nu$, where $t^\mu$ is timelike, we can always boost to a frame in which the speed of the graviton is isotropic, and the dispersion relation has the form
\be \label{gooddis}
\omega^2 + v^2 \vec{k}\cdot\vec{k} = 0,
\ee
where the propagation speed is different from the speed of light.
$H_\mn = t_\mu t_\nu$ thus defines a preferred rest frame, in which $t_\mu = (1,0,0,0)$.


\subsection{Motion of test particles}

We now want to investigate how the modified dispersion affects the motion of test particles in the presence of the transverse Goldstone modes. Consider nearby particles with separation vector $S^\mu$. The geodesic deviation equation of the test particles is
\be \label{geodesic}
\frac{D^2}{d\tau^2}S^\mu = {R^\mu}_{\nu\rs}U^\nu U^\rho S^\sigma,
\ee
where $R_{\mn\rs}$ is the Riemann tensor, $\tau$ is the proper time, and $U^\mu$ is the four-velocity of the test particles. The notation $\frac{D}{d\tau} = \frac{dx^\mu}{d\tau} \nabla_\mu$ denotes the directional covariant derivative. 

To first order, we can set $U^\mu = (1,0,0,0)$. Likewise, we can replace the Riemann tensor by its linearized version and the proper time $\tau$ by t. Eq.\eqref{geodesic} then becomes
\be \label{geofirstorder}
\frac{\pa^2}{\pa t^2}S^\mu = {{R^{(1)}}^\mu}_{00\sigma} S^\sigma,
\ee
where  
\be
R^{(1)}_{\mn \rs} = \frac{1}{2}(\pa_\rho \pa_\nu h_{\mu\sigma} + \pa_\sigma\pa_\mu h_{\nu\rho} - \pa_\sigma \pa_\nu h_{\mu\rho} - \pa_\rho \pa_\mu h_{\nu\sigma}).
\ee

For simplicity, we assume that the transverse modes propagate in the $z$ direction, so that $k^\mu = (\omega,0,0,k)$. Note that $\omega \neq k$, since the dispersion is no longer $k^\mu k_\mu=0$. As is shown in Appendix B \eqref{mostgeneral}, the polarization tensor of the two transverse modes is
\be \label{generalpol}
p_\mn = \left( \begin{array}{cccc} 
p_{00} & p_{10} & p_{20} & -p_{00} \\
p_{10} & h_+ & h_\times & -p_{10} \\
p_{20} & h_\times & -h_+ & -p_{20} \\
-p_{00} & -p_{10} & -p_{20} & p_{00} \end{array}\right),
\ee
where $h_\mn = p_\mn e^{ik_\alpha x^\alpha}$. The constants $p_{00},
p_{10},$ and $p_{20}$ can be determined by imposing the cardinal gauge
conditions\footnote{In Appendix B, we give an explicit formula for
  $p_{00}, p_{10},$ and $p_{20}$ in terms of components of
  $H_\mn$. The constants as they appear in \eqref{generalpol} are of the form $k_{(\mu}\xi_{|\nu)}$. They are therefore just gauge modes, so they are not physically observable if the theory is diffeomorphism invariant (as in general relativity). In Goldstone gravity, however, diffeomorphism invariance is broken, so the cardinal gauge mode components $p_{01}$ and $p_{02}$ in \eqref{deflection1} and \eqref{deflection2} can actually effect the motion of test particles, once radiative corrections are included.}
Because we do not start from a diffeomorphism-invariant formulation, we do not have the gauge freedom to set these coefficients to zero.

In Fourier space, Eq.\eqref{geofirstorder} becomes
\bea \nonumber
\omega^2 \delta S^\mu &=& \frac{1}{2}(\omega^2 {p^\mu}_\sigma + k_\sigma k^\mu p_{00} + \\ \label{geofourier}
&& k_\sigma \omega {p^\mu}_0 + \omega k^\mu p_{0\sigma})S^\sigma(0),
\eea
where $S^\mu(x^\mu) = S^\mu(0) + \delta S^\mu(x^\mu)$, and $S^\mu(0) = S^\mu(t=0,\vec{x}=\vec{0})$ is the initial position of the test particle.

With $h_{\mn} \propto e^{ik_\mu x^\mu}$, the zeroth component of Eq.\eqref{geofourier} reads
\bea \nonumber
\omega^2 \delta S^0 &=& \frac{1}{2}(\omega^2 {p^0}_\sigma + k_\sigma k^0 p_{00} 
\\ \nonumber
&& + k_\sigma \omega {p^0}_0 + \omega k^0 p_{0\sigma})S^\sigma(0) 
\\ \label{geozero}
&=& 0,
\eea
which is identically zero. There is no deflection in the time direction, as expected. 

For $\mu = 1$, we have
\bea \nonumber
\omega^2 \delta S^1 &=& \frac{1}{2}(\omega^2 h_{+} S^1(0) + \omega^2 h_{\times} S^2(0) \\
&& + (-\omega^2+k\omega)p_{01}S^3(0)).
\eea
If the dispersion relation is simply $k^\mu k_\mu = 0$, the last term is zero. However, with the modification $c H^\mn k_\mu k_\nu$ in the dispersion, $\omega \neq k$, and 
\bea \nonumber
\delta S^1 &=& \frac{1}{2}\big[h_{+} S^1(0)+h_{\times} S^2(0) \\ \label{deflection1}
&& -\frac{c_2}{2}(H_{33}+H_{00}+2H_{03})p_{01}S^3(0)\big].
\eea

Following the same procedure, the $\mu =2$ equation reads
\bea \nonumber
\delta S^2 &=& \frac{1}{2}(h_{\times} S^1(0)+h_{+} S^2(0) \\ \label{deflection2}
&& -\frac{c_2}{2}(H_{33}+H_{00}+2H_{03})p_{02}S^3(0)).
\eea
The first two terms in \eqref{deflection1} and \eqref{deflection2} correspond to the usual $+$ and $\times$ polarizations. However, both $\delta S^1$ and $\delta S^2$ are now also functions of the longitudinal separation $S^3(0)$. 

Similar to Eq.\eqref{geozero}, the $\mu=3$ equation is normally identically zero. However, because of the modified dispersion, we have 
\be
\delta S^3 = -\frac{c_2}{2}(H_{00}+H_{33}+2H_{03})(p_{01}S^1(0) + p_{02} S^2(0)).
\ee
Thus, the test particles will also undergo longitudinal oscillations. Notice that the amplitude of the oscillation is a function of the transverse position of the test particles. Hence the motion is not uniform along $z$. 

Similar to the graviton in general relativity, the two transverse modes have two polarizations (conveniently labelled $+$ and $\times$ here). The novel feature is that now both polarizations are accompanied by transverse oscillations that depend on longitudinal separation, and longitudinal oscillations that depend on transverse separation. 

\subsection{Experimental constraints}

If the speed of gravity $v_{graviton}$ is less than the speed of light, ultra-high energy cosmic rays will be able to emit `gravi-Cherenkov radiation'. This is analogous to the way in which a light source emits Cherenkov radiation in a medium. The fact that we observe ultra-high-energy cosmic rays puts a limit on the effectiveness of gravi-Cherenkov radiation, thereby placing a stringent lower bound on the propagation speed of the Goldstone modes (if they are to be interpreted as the graviton). We will use this to constrain the magnitude of the correction term $c_2 H^{\mn}k_\mu k_\nu$ in the graviton dispersion relation \eqref{moddis}. 

In \cite{Moore:2001bv}, it was found that, if gravi-Cherenkov radiation occurs, the maximum travelling time of a cosmic ray is
\be
t_{max} = \frac{M^2_{Pl}}{(n-1)^2 p^3},
\ee
where $p$ is the final momentum (when detected on Earth) and $n=v_{cosmic}/v_{graviton}$ is the refractive index. 

Using estimates in \cite{Moore:2001bv}, this translates to
\be
n-1 \approx \frac{c_2}{2}n_\mu H^{\mn} n_\nu < 2 \times 10^{-15}.
\ee
The speed of the Goldstone graviton can thus only be very slightly less than the speed of light.  
 
\subsection{Corrections to the energy-momentum tensor}

The correction to the dispersion relation also has an effect on the energy-momentum tensor of the transverse Goldstone modes. 

We define the energy-momentum tensor to be
\be \label{emtensor}
t_{\mn} = -\frac{1}{8\pi G}\left(R^{(2)}_\mn[h^{(1)}]-\frac{1}{2}\eta^{\rs} R_{\rs}^{(2)}[h^{(1)}]\eta_\mn \right),
\ee
where $R_\mn^{(i)}[h^{(j)}]$ is the parts of the expanded Ricci tensor that are $i^{th}$-order in the metric perturbation, while $h^{j}$ is the $j^{th}$-order expansion of the field $h_\mn$. Hence, $R_\mn^{(i)}[h^{(j)}]$ is of order $h^{(i\times j)}$. 

As $t_\mn$ is not diffeomorphism invariant, we should average over several wavelengths to obtain a reasonable measure of the energy-momentum. Imposing the cardinal conditions obeyed by the transverse Goldstone modes, Eq.\eqref{emtensor} simplifies to
\be
t^{(0)}_\mn = \frac{1}{64\pi G}k_\mu k_\nu \epsilon^{(trans)}_\rs \epsilon_{(trans)}^\rs. 
\ee

With the modification to the dispersion relation of the gravitons,
$k_\mu$ changes as $ k_\mu \rightarrow k_\mu + \frac{c_2}{2} H_\mn k^\nu$ up to first order. The energy-momentum tensor \eqref{emtensor} becomes
\be
t_\mn = t^{(0)}_\mn + \frac{c_2 \pi}{16G}(h_+^2 + h_\times^2) H_{\mu\alpha}k^\alpha H_{\nu\beta}k^\beta.
\ee
The flux of energy and momentum carried by the transverse Goldstone modes are therefore anisotropic, depending on $H_\mn$. This makes sense, as the modes propagate at different phase velocities in different directions. 

It has been estimated that the energy flux due to a typical supernova explosion at cosmological distances is approximately $10^{-19} erg/cm^2/s$. Given the experimental constraint from gravi-Cherenkov radiation on the size of $c_2 H_\mn$, the corrections are undetectable with current technologies. 

\section{Vevs that do not break all six generators}

\subsection{Gravitons are not Necessarily Goldstone}

For vector fields, an expectation value along with the Maxwell kinetic term naturally leads to 
photon-like Goldstone modes, regardless of the form of the vev.  We start out with four degrees of freedom in the vector $A_\mu$. The direction parallel to the vev is a massive mode, while the three orthogonal directions are the massless Goldstone excitations. We can further form two linear combinations of the Goldstone modes, such that they are transverse and obey the dispersion relation $k^\mu k_\mu = 0$. The longitudinal mode does not propagate.  

A similar story holds in the graviton case, as long as all six generators of the Lorentz group are broken, giving rise to six Goldstone bosons. (See Table~1 for a comparison with the photon case.) In this case, diffeomorphism invariance is also completely broken, and the counting proceeds analogously.  We start with ten degrees of freedom in $h_\mn$. The four cardinal gauge conditions define four `directions' along which the massive modes live. This leaves six degrees of freedom for the six Goldstone bosons. Imposing the four transverse conditions $k^\mu h_\mn = 0$ leaves us with two linear combinations that obey the dispersion relation $k^\mu k_\mu = 0$. The remaining four longitudinal modes are auxiliary and do not propagate.

This particularly straightforward case is the one that we have been focusing on so far. In this section, we will explore what happens when not all six generators are broken by the vev. In this case, there can be residual diffeomorphism invariance in the theory.  The Lorentz-violating theory might still contain two massless modes to be interpreted as the graviton, which now originate from diffemorphism invariance rather than Lorentz violation (so they are more like the gravitons in general relativity). This can never happen in the photon case, because the vev always completely breaks gauge invariance. 

\subsection{An Example: Three Goldstone Bosons Only}

We now wish to examine in detail a theory whose vev gives rise to three Goldstone modes only. Consider the Lagrangian
\bea \nonumber
L &=& \frac{1}{2}[(\pa_\mu \tih^{\mu\nu})(\pa_\nu \tih)-(\pa_\mu \tih^{\rho\sigma})(\pa_\rho {\tih^{\mu}}_{\sigma})
\\ \nonumber
&& +\frac{1}{2}\eta^{\mu\nu}(\pa_\mu \tih^{\rho\sigma})(\pa_\nu \tih_{\rho\sigma}) -\frac{1}{2}\eta^{\mu\nu}(\pa_\mu \tih)(\pa_\nu \tih)]
\\ \label{lagsimplevev}
&& + \lambda(\tih^{\mu\nu} \tih_{\mu\nu} - m^2),
\eea
where, for simplicity, we choose the potential to be a Lagrange multiplier instead of a smooth potential. This fixes the length of $\tih_\mn = H_\mn + h_\mn$. The corresponding equations of motion are
\bea\label{eom}
Q_{\mu\nu\rho\sigma}G^{\rho\sigma} = 0,
\eea
where 
\bea \nonumber
G_{\mu\nu} &=& \frac{1}{2}(\pa_\sigma \pa_\nu {h^\sigma}_\mu + \pa_\sigma \pa_\mu {h^\sigma}_\nu - \pa_\mu\pa_\nu h 
\\
&& - \Box h_{\mu\nu} - \eta_{\mu\nu}\pa_\rho \pa_\lambda h^{\rho\lambda} + \eta_{\mu\nu} \Box h)
\eea
is the usual linearized Einstein tensor, and $Q_{\mu\nu\rho\sigma} = \eta_{\mu\rho}\eta_{\nu\sigma}- \frac{1}{m^2}H_\mn H_\rs$ is a projection operator. Thus, \eqref{eom} is essentially Einstein's equations projected onto the hypersurface orthogonal to $H^{\mu\nu}$. Note that we do not consider radiative corrections in this section. 

Since the equations are linear, it is more convenient to switch to Fourier space ($\pa_\mu \rightarrow i k_\mu$), turning the differential equations into algebraic ones, which can then be written as a $9 \times 9$ matrix equation.
Assume that $m^2>0$ in \eqref{lagsimplevev}, one possible vev that minimizes the potential is 
\be \label{ttvev}
H_{\mn}= \left( \begin{array}{cccc} 
1 & 0 & 0 & 0 \\
0 & 0 & 0 & 0 \\
0 & 0 & 0 & 0 \\
0 & 0 & 0 & 0 \end{array}\right),
\ee
which leads to three Goldstone modes (three boosts):
\be \label{badgoldstone}
h^{(Goldstone)}_{\mn} = \left( \begin{array}{cccc} 
0 & -\beta_1 & -\beta_2 & -\beta_3 \\
-\beta_1 & 0 & 0 & 0 \\
-\beta_2 & 0 & 0 & 0 \\
-\beta_3 & 0 & 0 & 0 \end{array}\right).
\ee
 
As we demonstrate in Appendix B (where we give the most general polarization tensor of a graviton propagating in the $z$ direction), it is impossible for a graviton to have all vanishing spatial components. Thus, no linear combinations of these three Goldstone modes in \eqref{badgoldstone} can possibly behave like the graviton. 

Nonetheless, the theory does contain two massless degrees of freedom, as we now demonstrate by directly solving the equations of motion. The first-order fixed-norm constraint $H_{\mu\nu}h^{\mu\nu}=0$ (essentially the second cardinal gauge condition) implies that $h_{00} = 0$. The linearized equations of motion in momentum space are then
\begin{widetext}
\be \label{mtxeqn}
\left( \begin{array}{ccccccccc} 
k_2^2 + k_3^2 & -k_1 k_2 & -k_1 k_3 & 0 & k_0 k_2 & k_0 k_3 & -2k_0 k_1 & 0 & -2k_0 k_1 \\
-k_1 k_2 & k_1^2 + k_3^2 & -k_2 k_3 & -2k_0 k_2 & k_0 k_1 & 0 & 0 & k_0 k_3 & -2k_0 k_2 \\
-k_1 k_3 & -k_2 k_3 & k_1^2 + k_2^2 & -2k_0 k_3 & 0 & k_0 k_1 & -2k_0 k_3 & k_0 k_2 & 0 \\
0 & -2k_0 k_2 & -2k_0 k_3 & 0 & 0 & 0 & 2(-k_0^2 +k_3^2) & -2k_2 k_3 & 2(-k_0^2 +k_2^2) \\
k_0 k_2 & k_0 k_1 & 0 & 0 & k_0^2 -k_3^2 & k_2 k_3 & 0 & k_1 k_3 & -2k_1 k_2 \\
k_0 k_3 & 0 & k_0 k_1 & 0 & k_2 k_3 & k_0^2 -k_1^2 & -2k_1 k_3 & k_1 k_2 & 0 \\
-2k_0 k_1 & 0 & -2k_0 k_3 & 2(-k_0^2 +k_3^2) & 0 & -2k_1 k_3 & 0 & 0 & 2(-k_0^2 +k_1^2) \\
0 & k_0 k_3 & k_0 k_2 & -2k_2 k_3 & k_1 k_3 & k_1 k_2 & 0 & k_0^2 -k_1^2 & 0 \\
-2k_0 k_1 & -2k_0 k_2 & 0 & 2(-k_0^2 +k_2^2) & -2k_1 k_2 & 0 & 2(-k_0^2 +k_1^2) & 0 & 0
\end{array}\right)
\left( \begin{array}{c}
h_{01} \\ h_{02} \\ h_{03} \\ h_{11}/2 \\ h_{12} \\ h_{13} \\ h_{22}/2 \\ h_{23} \\ h_{33}/2
\end{array}\right) = 0.
\ee
\end{widetext}
Without loss of generality (since rotational invariance is preserved), we align axes such that $k^\mu = (\omega,0,0,k)$. The equations of motion \eqref{mtxeqn} have three zero eigenvalues, which is consistent with the fact that there are three residual gauge degrees of freedom. Meanwhile, there are two eigenvalues $\omega^2 - k^2$, and setting them to zero yields the dispersion relation $-\omega^2 + k^2 = k^\mu k_\mu = 0$. The corresponding eigenvectors have polarization tensors
\be
p_\mn = \left( \begin{array}{cccc} 
0 & 0 & 0 & 0 \\
0 & 1 & 0 & 0 \\
0 & 0 & -1 & 0 \\
0 & 0 & 0 & 0 \end{array}\right),
\ee
and
\be
p_\mn = \left( \begin{array}{cccc} 
0 & 0 & 0 & 0 \\
0 & 0 & 1 & 0 \\
0 & 1 & 0 & 0 \\
0 & 0 & 0 & 0 \end{array}\right).
\ee
These are exactly the $+$ and $\times$ polarizations of a graviton propagating in the $z$ direction in general relativity. Thus, the theory does contain two massless gravitons, but they do not arise as Goldstone bosons of spontaneous Lorentz violation. 

\begin{widetext}
\begin{center}
\begin{table}[ht]
\caption{Comparison between Goldstone Photons and \\Gravitons}
\centering
\begin{tabular}{c c c}
\hline\hline
& Photon & Graviton \\
\hline
Number of Goldstone Modes & $3$ & $6$ \\ \hline
Equivalent Gauge Conditions & Temporal or Axial & Cardinal \\ \hline
Number of Gauge Conditions/Massive Modes & $1$ & $4$ \\ \hline
Number of Transverse Modes & $2$ & $2$ \\ \hline
Number of Longitudinal Modes & $1$ & $4$ \\ \hline
Kinetic Term & Maxwell & Einstein-Hilbert \\ \hline\hline
\label{comparepg}
\end{tabular}
\end{table}
\end{center}
\end{widetext}

The origin of these massless excitations are more appropriately
associated with residual diffeomorphism invariance. With the chosen
ground state \eqref{ttvev}, the Lagrangian remains invariant under the
transformation $h_{\mn} \rightarrow h_{\mn} + \pa_\mu \xi_\nu +
\pa_\nu \xi_\mu$ for three independent functions $\xi_i$
(corresponding to the three zero eigenvalues of the equations of
motion). This guarantees the lack of mass terms for the components
$h_+$ and $h_\times$ in the Lagrangian.  

Furthermore, the simple vev \eqref{ttvev} gives only two, rather than four, cardinal gauge conditions. There are thus fewer massive `directions' in spacetime. Of the four conditions in \eqref{gaugecond}, only two are independent. Since $H_\mn \propto H_{\mu\rho}{H^\rho}_\nu \propto H_{\mu\rho} H^\rs H_{\sigma \nu}$, the last two gauge conditions in \eqref{gaugecond} are equivalent to the second. There are thus two, rather than four, massive modes. 

Let's compare this theory with the one that we have been considering in earlier sections. Before, the vev broke both Lorentz invariance and diffeomorphism completely. There are four cardinal gauge conditions, which implies that there are four massive modes. The remaining six degrees of freedom correspond to the six broken generators of the Lorentz group. Two linear combinations of the six propagate, while the remaining four are auxiliary. Together, they add up to the ten degrees of freedom in $h_\mn$.

In contrast, the theory that we consider in this section has a vev that breaks diffeomorphism invariance only partially. There are three remaining pure gauge modes. Because the vev preserves rotational invariance, only the three boost generators are broken, resulting in three Goldstone modes; none of them propagates, however. There are also only two massive modes, as the vev gives rise to only two independent cardinal gauge conditions. Together with the remaining two massless excitations that are identical to the graviton in general relativity, they account for the ten degrees of freedom that we started with in $h_\mn$. 

The possibilities are thus far richer in the graviton case than the photon case. In the former, there are three possibilities: the vev can break three, five, or six generators of the Lorentz group (We only discuss the first and the last case in this paper.) When there are fewer than six Goldstone bosons, it is possible that the theory has residual diffeomorphism invariance, which can also result in massless excitations with the right properties to be interpreted as the graviton.

\section{Conclusions}

Recently, Kostelecky and Potting \cite{Kostelecky:2009zr} examined in detail a scenario in which a symmetric two-index tensor acquires a vev via a potential. Two linear combinations of the six resulting Goldstone modes are dynamical and have properties identical to those of the graviton in general relativity. Because they originate in spontaneous symmetry breaking, this would provide a natural explanation for why the graviton is massless, without the need to invoke gauge invariance. 

It was pointed out in \cite{Kraus:2002sa} that, if we view the theory as an effective field theory, we should integrate out the massive modes, which would generate an infinite number of radiative correction terms in the low-energy effective Lagrangian. These terms are covariant in form, but involve the vev $H_\mn$, thereby inducing additional Lorentz-violating effects. In this paper, we examined the phenomenological properties of a subset of these radiative correction terms. In particular, we showed that they modify the dispersion relation of the two dynamical degrees of freedom, which becomes
\be\label{moddis2}
k_\mu k^\mu - c_2 H^{\mn}k_\mu k_\nu = 0.
\ee
This implies that the phase velocity of the dynamical modes is in general anisotropic. Another interesting consequence of the modified dispersion \eqref{moddis2} is that test particles in their vicinity would be deflected differently from those near the graviton in general relativity. They would undergo both transverse and longitudinal oscillations that depend on the longitudinal and transverse separation, respectively.

We also investigated the relationship between different forms of the vev $H_\mn$ and the corresponding Goldstone modes.  Unlike in the photon case, for gravity there exist vevs for which there are not enough Goldstone modes to construct the conventional graviton -- the gravitons may exist, but not as broken-symmetry generators acting on the vev.

Our analysis of the radiative-correction terms is by no means complete. For one thing, we have left out their effects on the four remaining Goldstone modes that become dynamical when they are present. Also, we only discussed terms that are linear in $H_\mn$ and ignored higher order corrections, which we believe to be sub-dominant, since Lorentz invariance has been verified to great accuracy at low energies. However, it is conceivable that the higher order corrections would lead to interesting effects in addition to those that are discussed in this paper, so they certainly merit further investigation. Finally, it would also be worthwhile to check whether the presence of the radiative corrections destabilize the theory. 

\section{Acknowledgements}
We are grateful to Arthur Lipstein for useful discussions. This work was supported in part by the U.S. Department of Energy and by the Gordon and Betty Moore Foundation.  IKW acknowledges financial
  support from the Research Council of Norway. 

\appendix
\section{Polarizations of Goldstone Modes}
We enumerate here the Goldstone modes that arise when a symmetric two-index tensor acquires various forms of vacuum expectation values. Linearity implies that the Goldstone mode corresponding to a general vev is a superposition of these modes.  

\subsection{Time-Time: $\mn = 00$}
Let's first consider the case where only the $00$ component of $H_{\mn}$ does not vanish. In that case, the three boost generators are broken, and we therefore have three Goldstone modes. 
\begin{scriptsize}
\be
H_{\mn}= \left( \begin{array}{cccc} 
1 & 0 & 0 & 0 \\
0 & 0 & 0 & 0 \\
0 & 0 & 0 & 0 \\
0 & 0 & 0 & 0 \end{array}\right) \rightarrow h_{\mn} = \left( \begin{array}{cccc} 
0 & -\beta_1 & -\beta_2 & -\beta_3 \\
-\beta_1 & 0 & 0 & 0 \\
-\beta_2 & 0 & 0 & 0 \\
-\beta_3 & 0 & 0 & 0 \end{array}\right)
\ee
\end{scriptsize}
Obviously, this choice of the vacuum expectation value preserves rotational invariance. Hence, none of the $\theta$ modes is excited. 

\subsection{Time-Space: $\mn = 0i$ or $i0$}
Now consider the case where one of the $0i$ components is non-zero. This breaks all three boosts, but only two of the three rotation generators. There are thus five Goldstone modes. 
\begin{scriptsize}
\be
H_{\mn}= \left( \begin{array}{cccc} 
0 & 1 & 0 & 0 \\
1 & 0 & 0 & 0 \\
0 & 0 & 0 & 0 \\
0 & 0 & 0 & 0 \end{array}\right) \rightarrow h_{\mn} = \left( \begin{array}{cccc} 
-2\beta_1 & 0 & \theta_3 & -\theta_2 \\
0 & -2\beta_1 & -\beta_2 & -\beta_3 \\
\theta_3 & -\beta_2 & 0 & 0 \\
-\theta_2 & -\beta_3 & 0 & 0 \end{array}\right).
\ee

\be
H_{\mn}= \left( \begin{array}{cccc} 
0 & 0 & 1 & 0 \\
0 & 0 & 0 & 0 \\
1 & 0 & 0 & 0 \\
0 & 0 & 0 & 0 \end{array}\right) \rightarrow h_{\mn} = \left( \begin{array}{cccc} 
-2\beta_2 & -\theta_3 & 0 & \theta_1 \\
-\theta_3 & 0 & -\beta_1 & 0 \\
0 & -\beta_1 & -2\beta_2 & -\beta_3 \\
\theta_1 & 0 & -\beta_3 & 0 \end{array}\right).
\ee

\be
H_{\mn}= \left( \begin{array}{cccc} 
0 & 0 & 0 & 1 \\
0 & 0 & 0 & 0 \\
0 & 0 & 0 & 0 \\
1 & 0 & 0 & 0 \end{array}\right) \rightarrow h_{\mn} = \left( \begin{array}{cccc} 
-2\beta_3 & \theta_2 & -\theta_1 & 0 \\
\theta_2 & 0 & 0 & -\beta_1 \\
-\theta_1 & 0 & 0 & -\beta_2 \\
0 & -\beta_1 & -\beta_2 & -2\beta_3 \end{array}\right).
\ee
\end{scriptsize}

\subsection{Diagonal Space-Space: $\mn = ii$}
\label{diag}
Now consider the case where one of the diagonal spatial elements does not vanish. This breaks one of the three boosts, and two of the rotations.
\begin{scriptsize}
\be
H_{\mn}= \left( \begin{array}{cccc} 
0 & 0 & 0 & 0 \\
0 & 1 & 0 & 0 \\
0 & 0 & 0 & 0 \\
0 & 0 & 0 & 0 \end{array}\right) \rightarrow h_{\mn} = \left( \begin{array}{cccc} 
0 & -\beta_1 & 0 & 0 \\
-\beta_1 & 0 & \theta_3 & -\theta_2 \\
0 & \theta_3 & 0 & 0 \\
0 & -\theta_2 & 0 & 0 \end{array}\right).
\ee

\be
H_{\mn}= \left( \begin{array}{cccc} 
0 & 0 & 0 & 0 \\
0 & 0 & 0 & 0 \\
0 & 0 & 1 & 0 \\
0 & 0 & 0 & 0 \end{array}\right) \rightarrow h_{\mn} = \left( \begin{array}{cccc} 
0 & 0 & -\beta_2 & 0 \\
0 & 0 & -\theta_3 & 0 \\
-\beta_2 & -\theta_3 & 0 & \theta_1 \\
0 & 0 & \theta_1 & 0 \end{array}\right).
\ee

\be
H_{\mn}= \left( \begin{array}{cccc} 
0 & 0 & 0 & 0 \\
0 & 0 & 0 & 0 \\
0 & 0 & 0 & 0 \\
0 & 0 & 0 & 1 \end{array}\right) \rightarrow h_{\mn} = \left( \begin{array}{cccc} 
0 & 0 & 0 & -\beta_3 \\
0 & 0 & 0 & \theta_2 \\
0 & 0 & 0 & -\theta_1 \\
-\beta_3 & \theta_2 & -\theta_1 & 0 \end{array}\right).
\ee
\end{scriptsize}

\subsection{Off-Diagonal Space-Space: $\mn = ij$} 
\label{offdiag}
Finally, we consider the case in which one of the off-diagonal spatial components is non-zero. This breaks two boosts and all rotations. 
\begin{scriptsize}
\be
H_{\mn}= \left( \begin{array}{cccc} 
0 & 0 & 0 & 0 \\
0 & 0 & 1 & 0 \\
0 & 1 & 0 & 0 \\
0 & 0 & 0 & 0 \end{array}\right) \rightarrow h_{\mn} = \left( \begin{array}{cccc} 
0 & -\beta_2 & -\beta_1 & 0 \\
-\beta_2 & -2\theta_3 & 0 & \theta_1 \\
-\beta_1 & 0 & 2\theta_3 & -\theta_2 \\
0 & \theta_1 & -\theta_2 & 0 \end{array}\right).
\ee

\be
H_{\mn}= \left( \begin{array}{cccc} 
0 & 0 & 0 & 0 \\
0 & 0 & 0 & 1 \\
0 & 0 & 0 & 0 \\
0 & 1 & 0 & 0 \end{array}\right) \rightarrow h_{\mn} = \left( \begin{array}{cccc} 
0 & -\beta_3 & 0 & -\beta_1 \\
-\beta_3 & 2\theta_2 & -\theta_1 & 0 \\
0 & -\theta_1 & 0 & \theta_3 \\
-\beta_1 & 0 & \theta_3 & -2\theta_2 \end{array}\right).
\ee

\be
H_{\mn}= \left( \begin{array}{cccc} 
0 & 0 & 0 & 0 \\
0 & 0 & 0 & 0 \\
0 & 0 & 0 & 1 \\
0 & 0 & 1 & 0 \end{array}\right) \rightarrow h_{\mn} = \left( \begin{array}{cccc} 
0 & 0 & -\beta_3 & -\beta_2 \\
0 & 0 & \theta_2 & -\theta_3 \\
-\beta_3 & \theta_2 & -2\theta_1 & 0 \\
-\beta_2 & -\theta_3 & 0 & 2\theta_1 \end{array}\right).
\ee
\end{scriptsize}

Notice that not all ten modes are independent. We can, for example, perform a rotation to diagonalize the three modes in \ref{offdiag}, so that they become a linear combination of the modes in \ref{diag}. 

\section{Proof that gravitons can be Goldstone bosons}
We present here a proof that when all six generators are broken, two linear combinations of the resulting six Goldstone bosons have properties that agree with those of the graviton at lowest order.\footnote{During the preparation of this manuscript, we became aware of the recent work by Kostelecky and Potting \cite{Kostelecky:2009zr}, in which they gave a proof that a version of this Lorentz-violating theory of gravity is identical to linearized gravity in the cardinal gauge.} The propagating Goldstone modes obey the dispersion relation $k^\mu k_\mu = 0$, the transverse conditions $k^\mu h_\mn = 0$, and the four cardinal gauge conditions.  

First consider the most general vacuum expectation value 
\be \label{complicatedvev}
H_\mn= \left( \begin{array}{cccc} 
d & e & f & g \\
e & a & h & i \\
f & h & b & j \\
g & i & j & c \end{array}\right),
\ee
where the ten constants $a$, $b$, $c$, $d$, $e$, $f$, $g$, $h$, $i$, $j$ are presumably determined by the potential V in \eqref{grlagrangian}. This choice of the vev might seem unnecessarily complicated (as it can be simplified by boosts and rotations). However, as will be shown below, Eq.\eqref{complicatedvev} will simplify our analysis later on.

This vacuum expectation value gives the following Goldstone excitations (see Appendix A):

\bea \label{h00}
h_{00} &=& -2e\beta_1 - 2f\beta_2 - 2g\beta_3
\\
h_{01} &=& -(a+d)\beta_1 - h\beta_2 - i\beta_3 + g\theta_2 - f\theta_3
\\
h_{02} &=& -h\beta_1 - (b+d)\beta_2 - j\beta_3 - g\theta_1 + e\theta_3
\\
h_{03} &=& -i\beta_1 - j\beta_2 - (c+d)\beta_3 + f\theta_1 - e\theta_2
\\
h_{11} &=& -2e\beta_1 + 2i\theta_2 - 2h\theta_3
\\
h_{22} &=& -2f\beta_2 - 2j\theta_1 + 2h\theta_3
\\
h_{33} &=& -2g\beta_3 + 2j\theta_1 - 2i\theta_2
\\
h_{12} &=& -f\beta_1 - e\beta_2 - i\theta_1 + j\theta_2 + (a-b)\theta_3
\\
h_{13} &=& -g\beta_1 - e\beta_3 + h\theta_1 + (c-a)\theta_2 - j\theta_3
\\ \label{h23}
h_{23} &=& -g\beta_2 - f\beta_3 + (b-c)\theta_1 - h\theta_2 + i\theta_3
\eea

We would now like to demonstrate that it is possible for the Goldstone modes resulting from a completely general vev to have a polarization tensor that agrees with that of a graviton (in GR) propagating in the $z$ direction in some gauge. In general relativity, we have the freedom to add to any solution of the linearized Einstein's equations the pure gauge mode $k_{(\mu|}\xi_{|\nu)}$. Therefore, the familiar $+$ and $\times$ polarizations in the transverse-traceless gauge,
\be \label{tthmnz}
h^{TT}_{\mn}= \left( \begin{array}{cccc} 
0 & 0 & 0 & 0 \\
0 & h_+ & h_\times & 0 \\
0 & h_\times & -h_+ & 0 \\
0 & 0 & 0 & 0 \end{array}\right) e^{ik_\alpha x^\alpha},
\ee 
are not the most general form that the graviton in general relativity can take. 

For a graviton propagating in the $z$ direction, we have $k^\mu=(\omega,0,0,\omega)$. If we set $\xi_\mu=\frac{1}{\omega}(-p_{00},-p_{01},-p_{02},-p_{03})$, the polarization $p^{(gauge)}_\mn$ of the most general gauge mode $h^{(gauge)}_\mn = p^{(gauge)}_\mn e^{ik_\alpha x^\alpha}$ can be written as
\be \label{gauge}
p^{(gauge)}_\mn = \left( \begin{array}{cccc} 
p_{00} & p_{01} & p_{02} & (p_{03}-p_{00})/2 \\
p_{01} & 0 & 0 & -p_{01} \\
p_{02} & 0 & 0 & -p_{02} \\
(p_{03}-p_{00})/2 & -p_{01} & -p_{02} & -p_{03} \end{array}\right),
\ee
where $p_{00}, p_{01}, p_{02},$ and $p_{03}$ are constants. Thus, the most general form that the graviton can assume in GR is the sum of \eqref{tthmnz} and \eqref{gauge}\footnote{Here, we are restricting ourselves to graviton solutions of the form $e^{ik_\alpha x^\alpha}$. If we relax this assumption, it is conceivable that there are other possible functional forms. This is analogous to electromagnetism in the axial gauge, in which $A_\mu \propto z e^{ik_\alpha x^\alpha}$ is needed to describe a plane-wave photon in the $z$ direction. Thus, the field becomes unbounded at spatial infinity, and it is questionable whether our effective theory is valid.}:
\be \label{mostgeneral}
h^{(general)}_\mn = \left( \begin{array}{cccc} 
p_{00} & p_{01} & p_{02} & -p_{00} \\
p_{01} & h_+ & h_\times & -p_{01} \\
p_{02} & h_\times & -h_+ & -p_{02} \\
-p_{00} & -p_{01} & -p_{02} & p_{00} \end{array}\right)e^{ik_\alpha x^\alpha}.
\ee
Note that because the Goldstone modes are all traceless, we have also set $p_{00} = -p_{03}$ above. We now want to see if the polarizations of the Goldstone bosons resulting from the most general vev \eqref{complicatedvev} can be matched onto \eqref{mostgeneral}.  

To match \eqref{complicatedvev} onto \eqref{mostgeneral}, we have to satisfy the following conditions:
\bea \nonumber
h_{00} &=& -h_{03} \\ \nonumber
h_{01} &=& -h_{31} \\ \nonumber
h_{02} &=& -h_{32} \\ \label{generalcond}
h_{00} &=& h_{33}.
\eea
These four conditions leave in the six Goldstone modes two degrees of freedom, exactly the right number to describe the graviton, which has two polarizations. 

At this point, it is convenient to define new fields by linearly combining the Goldstone modes:
\bea \nonumber
M_1 &=& -(h_{00}+h_{33})  
\\ \nonumber
&=& (2e+i)\beta_1 + (2f+j)\beta_2 +(2g+c+d)\beta_3
\\
&& - f\theta_1 + e\theta_2
\\ \nonumber
M_2 &=& -(h_{01}+h_{31})
\\ \nonumber
&=& (a+d+g)\beta_1 + h\beta_2 + (i+e)\beta_3
\\
&& - h\theta_1 - (g+c-a)\theta_2 +(f+j)\theta_3 
\\ \nonumber
M_3 &=& -(h_{02}+h_{32}) 
\\ \nonumber
&=& h\beta_1 +(b+d+g)\beta_2 + (j+f)\beta_3
\\
&& +(g+c-b)\theta_1 + h\theta_2 - (e+i)\theta_3
\\ \nonumber
M_4 &=& -h_{00}+h_{33} 
\\
&=& 2e\beta_1 + 2f\beta_2 + 2j\theta_1 -2i\theta_2
\\ \nonumber
M_5 &=& h_{11} \equiv h_+ 
\\
&=& -2e\beta_1 +2i\theta_2 - 2h\theta_3 
\\ \nonumber
M_6 &=& h_{12} \equiv h_\times
\\
&=& -f\beta_1 - e\beta_2 - i\theta_1 + j\theta_2 +(a-b)\theta_3.
\eea 
In this new basis, the physical degrees of freedom are made very transparent: $M_5$ and $M_6$ are the usual $+$ and $\times$ gravitons. The four conditions \eqref{generalcond} now become $M_1 = M_2 = M_3 = M_4 = 0$. 

These six linear equations relating the two bases can be written as a matrix equation
\be \label{matrixeqn}
\textbf{A} \vec{\zeta} = \vec{M},
\ee
where $\vec{\zeta} = (\beta_1, \beta_2, \beta_3, \theta_1, \theta_2, \theta_3)$ and $\vec{M} = (M_1, M_2, M_3, M_4, M_5, M_6)$ are the Goldstone modes in the original basis and new basis, respectively. This gives immediately the constraint det$(\textbf{A}) \neq 0$, since otherwise the matrix $\textbf{A}$ is singular and the new basis spanned by $\vec{M}$ is incomplete. 

To express $h_\mn$ in the new basis spanned by $\vec{M}$, we first invert Eq.\eqref{matrixeqn} to solve for $\vec{\zeta} = \textbf{A}^{-1} \vec{M}$, which can then be substituted into Eqs.\eqref{h00} - \eqref{h23}. 

\subsection{The Two Transverse Linear Combinations of the Six Goldstone Modes}
We now proceed to show that two linear combinations of the Goldstone modes ($M_5$ and $M_6$) obey the dispersion relation $k^\mu k_\mu = 0$ and are transverse to the momentum ($k^\mu h_\mn = 0$).

Setting all $M_i=0$ except for $M_5$, all the conditions in \eqref{generalcond} would be satisfied, and we have 
\be \label{plus}
h^{(5)}_\mn = \left( \begin{array}{cccc} 
c^{5}_{00} & c^{5}_{01} & c^{5}_{02} & -c^{5}_{00} \\
c^{5}_{01} & 1 & 0 & -c^{5}_{01} \\
c^{5}_{02} & 0 & -1 & -c^{5}_{02} \\
-c^{5}_{00} & -c^{5}_{01} & -c^{5}_{02} & c^{5}_{00} \end{array}\right)M_5,
\ee
which has exactly the form of \eqref{mostgeneral} if $h_\times = 0$. $M_5$ therefore corresponds to the $+$ polarization of the graviton. The constants $c^{5}_{ij}$ are computed straightforwardly using Eqs.\eqref{h00} - \eqref{h23}. 

Similarly, if we turn off all the $M_i$'s except $M_6$, all the conditions \eqref{generalcond} are satisfied, and the polarization tensor of the Goldstone mode $M_6$ becomes
\be \label{cross}
h^{(6)}_\mn = \left( \begin{array}{cccc} 
c^{6}_{00} & c^{6}_{01} & c^{6}_{02} & -c^{6}_{00} \\
c^{6}_{01} & 0 & 1 & -c^{6}_{01} \\
c^{6}_{02} & 1 & 0 & -c^{6}_{02} \\
-c^{6}_{00} & -c^{6}_{01} & -c^{6}_{02} & c^{6}_{00} \end{array}\right)M_6,
\ee
which agrees with \eqref{mostgeneral} if $h_+ = 0$, and therefore represents the $\times$ polarization. As before, the constants $c^{6}_{ij}$ are computed using Eqs.\eqref{h00} - \eqref{h23}. Note that because $M_5$ and $M_6$ are non-zero, it is in general impossible to set all $c^5_{ij}$ and $c^6_{ij} = 0$. That is, no choice of $H_\mn$ corresponds to the transverse-traceless gauge conventionally used to describe the graviton. 

Because the kinetic terms in the Lagrangian of our theory are those in the Einstein-Hilbert action, the equations of motion of these Goldstone modes (valid for all six modes $M_{1\rightarrow 6}$) to leading order are simply given by the linearized Einstein equation in vacuum
\be \label{gmunu}
\pa_\sigma \pa_\nu {h^\sigma}_\mu + \pa_\sigma \pa_\mu {h^\sigma}_\nu - \Box h_\mn - \eta_\mn \pa_\rho \pa_\lambda h^{\rho\sigma} = 0.
\ee

Substituting the $+$ mode, Eq.\eqref{plus}, into Eq.\eqref{gmunu} and
setting the 4-momentum to $k^\mu = (\omega,0,0,k)$ gives 
\bea \label{G00}
2G_{00} &=& 0
\\
2G_{01} &=& c^5_{01}k(\omega - k) = 0
\\
2G_{02} &=& c^5_{02}k(\omega - k) = 0
\\
2G_{03} &=& 0
\\
2G_{11} &=& (\omega^2 - k^2) - (\omega - k)^2 c^5_{00} = 0
\\
2G_{12} &=& 0
\\
2G_{13} &=& c^5_{12}\omega(k - \omega) = 0
\\
2G_{22} &=& -(\omega^2 - k^2) - (\omega - k)^2 c^5_{00} = 0
\\
2G_{23} &=& c^5_{23}\omega(k - \omega) = 0
\\ \label{G33}
2G_{33} &=& 0.
\eea
In general, $c^5_{ij}$ do not vanish and Eqs.\eqref{G00} - \eqref{G33} imply that $\omega = k$. That is, $h^{(5)}_\mn$ propagates along the $z$ direction at the speed of light, as expected. 

If instead we substitute the $\times$ mode (Eq.\eqref{cross}) into Eq.\eqref{gmunu} and again set the 4-momentum $k^\mu = (\omega,0,0,k)$, we obtain the same equations, except that now
\bea
2G_{11} &=& -(\omega - k)^2 c^6_{00} = 0 
\\
2G_{12} &=& (\omega^2 - k^2) = 0
\\
2G_{22} &=& -(\omega - k)^2 c^6_{00} = 0, 
\eea
and $c^5_{ij} \rightarrow c^6_{ij}$ in \eqref{G00} - \eqref{G33}. Clearly, the solution is still $\omega = k$. Thus, $h^6_\mn$ also propagates along $z$ at the speed of light. 

Finally, the fact that these modes are transverse can be shown by direct computation:
\bea \nonumber
k^\mu h^{(general)}_\mn &=& k^\mu(h^{TT}_\mn + p^{gauge}_\mn e^{ik_\alpha x^\alpha}) \\ \nonumber
&=& \frac{1}{2}k^\mu(k_\mu \xi_\nu + k_\nu \xi_\mu) \\ \nonumber
&=& \frac{1}{2}(k^2 \xi_\nu + k_\nu k^\mu \xi_\mu) \\
&=& 0,
\eea
since the graviton obeys $k^2=0$ and the gauge modes are traceless (i.e. $k_\mu \xi^\mu=0$). 

In summary, we have shown that there are two special linear combinations ($M_5$ and $M_6$) of the six Goldstone modes that have a polarization tensor identical to that of a graviton in general relativity; obey the normal dispersion relation $k^2=0$; and are transverse to the momentum $k^\mu$. 

\subsection{The Remaining Four Linear Combinations}
In this section, we demonstrate that the remaining four linear combinations do not propagate upon imposing the equations of motion.
The four remaining modes ($M_1$ to $M_4$) are given respectively by
\bea \label{mode1}
h^{(1)}_\mn = \left( \begin{array}{cccc} 
c^{1}_{00} & c^{1}_{01} & c^{1}_{02} & c^{1}_{03} \\
c^{1}_{01} & 0 & 0 & -c^{1}_{01} \\
c^{1}_{02} & 0 & 0 & -c^{1}_{02} \\
c^{1}_{03} & -c^{1}_{01} & -c^{1}_{02} & c^{1}_{00} \end{array}\right)M_1
\\ \label{mode2}
h^{(2)}_\mn = \left( \begin{array}{cccc} 
c^{2}_{00} & c^{2}_{01} & c^{2}_{02} & -c^{2}_{00} \\
c^{2}_{01} & 0 & 0 & c^{2}_{13} \\
c^{2}_{02} & 0 & 0 & -c^{2}_{02} \\
-c^{2}_{00} & c^{2}_{13} & -c^{2}_{02} & c^{2}_{00} \end{array}\right)M_2
\\ \label{mode3}
h^{(3)}_\mn = \left( \begin{array}{cccc} 
c^{3}_{00} & c^{3}_{01} & c^{3}_{02} & -c^{3}_{00} \\
c^{3}_{01} & 0 & 0 & -c^{3}_{01} \\
c^{3}_{02} & 0 & 0 & c^{3}_{23} \\
-c^{3}_{00} & -c^{3}_{01} & c^{3}_{23} & c^{3}_{00} \end{array}\right)M_3
\\ \label{mode4}
h^{(4)}_\mn = \left( \begin{array}{cccc} 
c^{4}_{00} & c^{4}_{01} & c^{4}_{02} & c^{4}_{03} \\
c^{4}_{01} & 0 & 0 & -c^{4}_{13} \\
c^{4}_{02} & 0 & 1 & c^{4}_{23} \\
c^{4}_{03} & c^{4}_{13} & c^{4}_{23} & c^{4}_{00}-c^{4}_{33} \end{array}\right)M_4,
\eea
where $c^1_{ij},c^2_{ij},c^3_{ij},c^4_{ij}$ are constants determined by Eqs.\eqref{h00} - \eqref{h23}. 

Again, using the linearized Einstein's equations, the mode $M_1$ \eqref{mode1} has the following equations of motion:
\bea
2G_{00} &=& -(c^1_{00}\omega^2 + 2k c^1_{03}\omega + c^1_{00} k^2) = 0
\\
2G_{01} &=& c^1_{01}k(\omega - k) = 0
\\
2G_{02} &=& c^2_{02}k(\omega - k) = 0
\\
2G_{03} &=& 0
\\
2G_{11} &=& c^1_{00}\omega^2 + 2k c^1_{03}\omega + c^1_{00} k^2 = 0
\\
2G_{12} &=& 0
\\
2G_{13} &=& c^1_{01}\omega(k - \omega) = 0
\\
2G_{22} &=& c^1_{00}\omega^2 + 2k c^1_{03}\omega + c^1_{00} k^2 = 0
\\
2G_{23} &=& c^1_{02}\omega(k - \omega) = 0
\\ 
2G_{33} &=& 0.
\eea
In general, the constants $c^1_{ij}$ do not vanish and the only way to satisfy all these conditions is to set $\omega = k = 0$. This mode therefore does not propagate. It is straightforward to repeat the analysis for the other three modes, and it can be shown that their equations of motion lead to $\omega = k = 0$.  

This analysis is thus in agreement with that by Kostelecky and Potting
\cite{Kostelecky:2009zr}: in this Lorentz-violating theory, only two linear combinations of the six Goldstone modes propagate and obey the dispersion relation $k_\mu k^\mu = 0$ and the transverse condition $k_\mu \epsilon^\mn = 0$. Also, because of the form \eqref{goldstone} of the Goldstone modes, the cardinal gauge conditions are all satisfied. The four remaining linear combinations do not propagate. Thus, at lowest order, the theory contains two propagating modes with properties identical to the graviton in linearized general relativity. 

\section{Proof of the necessity of breaking all six generators to get Goldstone gravitons}
We now discuss a systematic way of determining the number of Goldstone modes that result for a given vev. 
We construct a $10 \times 6$ matrix $\textbf{N}$ where each row corresponds to one of the ten components of $h_{\mu\nu}$, and each column corresponds to one of the six generators of the Lorentz group ($\theta_i$ and $\beta_i$, $i\in 1,2,3$). 
\begin{widetext}
\be
\textbf{N} = \left( \begin{array}{cccccc} 
-2H_{01} & -2H_{02} & -2H_{03} & 0 & 0 & 0 \\
-(H_{00}+h_{11}) & -H_{12} & -H_{13} & 0 & H_{03} & -H_{02} \\
-H_{12} & -(H_{00}+H_{22}) & -H_{23} & -H_{03} & 0 & H_{01} \\
-H_{13} & -H_{23} & -(H_{00}+H_{33}) & H_{02} & -H_{01} & 0 \\
-2H_{01} & 0 & 0 & 0 & 2H_{13} & -2H_{12} \\
-H_{02} & -H_{01} & 0 & -H_{13} & H_{23} & H_{11}-H_{22} \\
-H_{03} & 0 & -2H_{01} & H_{12} & H_{33}-H_{11} & -H_{23} \\
0 & -2H_{02} & 0 & -2H_{23} & 0 & 2H_{12} \\
0 & -H_{03} & -H_{02} & H_{22}-H_{33} & -H_{12} & H_{13} \\
0 & 0 & -2H_{03} & 2H_{23} & -2H_{13}& 0 \\ \end{array}\right),
\ee
\end{widetext}
The entries $\textbf{N}$ are the coefficients of the $\theta_i$ and $\beta_i$ in the ten components of $h_{\mn}$. The rank of this matrix is the number of Goldstone modes. The possible ranks of this matrix are three, five, and six. This is different in the vector case, in which the rank of the corresponding $4 \times 6$ matrix is always three, consistent with the fact that there are always three Goldstone modes.

We found in Appendix B that a necessary and sufficient condition for the theory to contain two linear combinations of the Goldstone modes is 
\be \label{detcond}
\mbox{det} (\textbf{A}) \neq 0,
\ee
which is equivalent to Rank$(\textbf{A}) = 6$. Since the rows of $\textbf{A}$ are just linear combinations of those of $\textbf{N}$, the rank of the former is necessarily less than or equal to the latter. Thus, for vevs that do not break all six generators, the number of Goldstone modes $<6$, implying that
\bea
&& \mbox{Rank}(\textbf{N})<6 
\\
&\Rightarrow& \mbox{Rank}(\textbf{A}) < 6 
\\
&\Leftrightarrow& \mbox{det}(\textbf{A}) = 0,
\eea
violating the condition \eqref{detcond}. This implies the lack of two linear combinations of the Goldstone modes that behave like the graviton in general relativity. However, as was discussed, it is still possible that the theory contains massless excitations that behave like the graviton; they are just not Goldstone in origin. 

\bibliography{goldstone-graviton}

\end{document}